\documentclass[pra,twocolumn,showpacs,floatfix,superscriptaddress,aps]{revtex4}
\usepackage{amsmath}
\usepackage{makeidx}
\usepackage{amssymb}
\usepackage{graphicx}

\usepackage[usenames]{color}
\usepackage[normalem]{ulem}

\usepackage{hyperref}
\usepackage[T1]{fontenc} 
\usepackage{soul}
\begin{document}
%%%%%%%%%%%%%%%%%%%%%%%%%%%%%%%%%%%%%%%%%%%%%%%%%%%%%%%%%%%%%%%%%%%%%%%%%%%%%%
%%%%                     Title and authors                                %%%%
%%%%%%%%%%%%%%%%%%%%%%%%%%%%%%%%%%%%%%%%%%%%%%%%%%%%%%%%%%%%%%%%%%%%%%%%%%%%%%

\title{Fractional-charge vortex dipoles in spinor Bose-Einstein condensates}

\author{Sandeep Gautam\footnote{sandeepgautam24@gmail.com}}
%\author{S. K. Adhikari\footnote{adhikari44@yahoo.com, 
%        URL  http://www.ift.unesp.br/users/adhikari}}
\affiliation{Instituto de F\'{\i}sica Te\'orica, Universidade Estadual
             Paulista - UNESP, \\ 01.140-070 S\~ao Paulo, S\~ao Paulo, Brazil}
      
%%%%%%%%%%%%%%%%%%%%%%%%%%%%%%%%%%%%%%%%%%%%%%%%%%%%%%%%%%%%%%%%%%%%%%%%%%%%%%
%%%%%%%%%%                    Abstract                             %%%%%%%%%%%
%%%%%%%%%%%%%%%%%%%%%%%%%%%%%%%%%%%%%%%%%%%%%%%%%%%%%%%%%%%%%%%%%%%%%%%%%%%%%%

\date{\today}
\begin{abstract} 
We theoretically and numerically investigate the generation of fractional-charge 
vortex dipoles in spinor condensates with non-zero magnetization. We find that 
in the antiferromagnetic phase of spin-1 and spin-2 and the cyclic phase of spin-2 condensate 
with non-zero magnetization coupling of the density (phonon) and a spin-excitation mode 
results in two critical speeds for vortex-antivortex pair creation in the condensate. 
As a result, a Gaussian obstacle potential moving across the antiferromagnetic spin-1 or spin-2 and cyclic spin-2 
spinor condensates with non-zero magnetization can lead to the creation of 
fractional-charge vortex dipoles. On the other hand, for zero magnetization,
the two modes get decoupled, which is illustrated by a single critical speed for
vortex-antivortex pair creation in the condensate due to the phonon excitation mode.
\end{abstract}

\maketitle

%%%%%%%%%%%%%%%%%%%%%%%%%%%%%%%%%%%%%%%%%%%%%%%%%%%%%%%%%%%%%%%%%%%%%%%%%%%%%%%
%%%%%                        Introduction                             %%%%%%%%%
%%%%%%%%%%%%%%%%%%%%%%%%%%%%%%%%%%%%%%%%%%%%%%%%%%%%%%%%%%%%%%%%%%%%%%%%%%%%%%%

\section{Introduction}
Quantized vortices are the topological excitations of superfluids like Bose-Einstein
condensates (BECs) \cite{Donnelly}. In contrast to scalar BECs, a wide variety of vortices
can exist in spinor BECs \cite{ueda, Stamper-Kurn,Masahito}. In scalar BECs, while traversing 
a closed loop around the vortex center, the gauge phase of the wave function
changes by an integer multiple $n$ of $2\pi$ to ensure the single-valuedness of
the wave function, where $n$ is the charge of the vortex. 
The circulation of the velocity field resulting in mass current is quantized to 
 $nh/M$ for these vortices, here $M$ is the mass of an atom \cite{Onsager, Fetter}. In contrast
to this, for the ferromagnetic phases of spin-1 and spin-2 BECs \cite{ueda,Masahito} 
with non-zero spin-expectation per particle, the circulation of the velocity field responsible for
the mass current can change continuously by changing
spin configurations or spin texture as in the case of coreless vortices
\cite{Mizushima-1,Mizushima-2,Mizushima-3,Leanhardt}.
In the case of antiferromagnetic phase of spin-1 \cite{Ohmi,Ho} and spin-2 and 
cyclic phase of spin-2 BECs \cite{Koashi,Ciobanu}, the gauge phase of the wave function
can change by a fractional multiple $n$ of $2\pi$ while traversing a closed loop around
the vortex center. This results in the emergence of a fractional-charge vortex with circulation of
the velocity field responsible for the mass current equal to a fractional multiple $n$ of $h/M$, e.g.
a half-quantum vortex or Alice vortex in an antiferromagnetic  spin-1 BEC \cite{Zhou,Seo}. 
In addition to these vortices, polar-core vortices can also exist in both ferromagnetic and 
antiferromagnetic phases of spin-1 BECs \cite{Mizushima-2,Mizushima-3,Isoshima,Sadler}; these vortices 
are characterized by component vortices with integer charges $\pm 1$ and 0 
in the spin components $m_f=\pm 1$ and 0, respectively \cite{ueda}. The nucleation of stable 
fractional-charge vortices in the cyclic spin-2 BEC under rotation has also been 
theoretically studied \cite{cyclic}.

A vortex dipole is a pair of vortices with opposite circulation and plays
crucial roles in the phenomena ranging from superfluid turbulence \cite{Tsubota,White}, 
the Berezinskii-Kosterlitz-Thouless (BKT) phase transition \cite{Berezinskii,Kosterlitz,Hadzibabic}, and 
phase transition dynamics via Kibble-Zurek mechanism \cite{Kibble-Zurek}, etc.
It has been shown by the numerical simulations that the superfluid flow past an obstacle 
above a critical velocity becomes dissipative via vortex dipoles generation \cite{Frisch}.
According to Landau criterion for the break down of the superfluidity, the critical speed 
for the dissipation to set in is equal to the minimum phase velocity, i.e. 
$v_c = {\rm min}[ \epsilon(p)/p]$, where $\epsilon(p)$
is the energy of an elementary excitation of momentum $p$ \cite{Landau}. The generation of vortex dipoles
by the superfluid flow past an obstacle above a critical speed has been confirmed 
experimentally \cite{Neely}. It has also been shown that the critical speed for vortex dipole generation with a 
penetrable Gaussian obstacle potential is a fraction of the local speed of sound \cite{Kwon}.

In this paper, we study the generation of fractional-charge vortex dipoles by a moving Gaussian
obstacle potential in a spin-1 antiferromagnetic BEC and spin-2 antiferromagnetic and cyclic BECs with non-zero magnetizations 
using a set of coupled mean-field Gross-Pitaevskii (GP) equations. We consider spinor BECs
in quasi-two-dimensional (quasi-2D) \cite{Salasnich} trapping potentials. The minimum energy states 
of spin-1 and spin-2 BECs with fractional-charge vortex dipoles have only two non-zero component spin 
states. This allows us to describe the system by a coupled set of two GP equations. We calculate 
the excitation spectrum and critical speeds of this binary system. We show that coupling of the 
phonon excitation mode and a spin excitation mode for non-zero magnetization results in two 
critical speeds for the vortex-antivortex pair (vortex dipole) creation. We exploit this coupling to
create fractional-charge vortex dipoles in spinor condensates with non-zero magnetization.
In case of spin-1 antiferromagnetic BEC and spin-2 antiferromagnetic and cyclic BECs with zero magnetization,
there is a single critical speed for vortex-antivortex pair creation due to the decoupling
of the aforementioned two modes. The result for flow of an antiferromagnetic spin-1 condensate
with zero magnetization across an obstacle are consistent with an earlier study \cite{Rodrigues}. 

 The paper 
is organized as follows. In Sec. \ref{sec-II}, we describe the coupled GP equations
for the spin-1 and spin-2 BECs. 
In Sec. \ref{3}, we classify different fractional-charge vortex dipoles 
of spin-1 and spin-2 BECs which we study in this paper.
Here, the interaction parameters of the equivalent binary
system under rotation are defined in terms of the interaction parameters of the  
full $(2f+1)$-component GP equation. 
In Sec. \ref{sec-IV}, we calculate the excitation spectrum of the uniform binary system.  
The numerical results for the generation fractional-charge vortex dipoles
in spin-1 and spin-2 BECs are presented in Sec. \ref{V}. 
We conclude the paper by giving a brief summary and discussion in Sec. \ref{VI}.

\section{GP equations for spin-1 and spin-2 BECs}
\label{sec-II}
We consider spinor BECs of $N$ atoms of mass $M$ each trapped
by quasi-2D traps with $\omega_z\gg\sqrt{\omega_x \omega_y}$, where $\omega_x,\omega_y,\omega_z$
are the confining trap frequencies in $x,y,z$ directions, respectively.
In mean-field approximation, such a spin-1 quasi-2D BEC is
described by the following set of three coupled two-dimensional 
GP equations for different spin components $m_f=\pm 1,0$ \cite{ueda,Ohmi,Ho}
 \begin{eqnarray}&&
\mu_{\pm 1} \psi_{\pm 1}(\mathbf r) =
 {\cal H}\psi_{\pm 1}(\mathbf r) 
\pm   c^{}_1F_z\psi_{\pm 1}(\mathbf r) 
+ \frac{c^{}_1}{\sqrt{2}} F_{\mp}\psi_0(\mathbf r),
 \label{gps-1}\\
&&\mu_0 \psi_0(\mathbf r) =
{\cal H}\psi_0(\mathbf r)  %  +c^{}_0n\phi_{0}(\mathbf r)
+ \frac{c_1}{\sqrt 2} [F_{-}\psi_{-1}(\mathbf r) 
+F_{+}\psi_{+1}(\mathbf r)]
\label{gps-2}, 
\end{eqnarray}
where ${\bf F}\equiv\{F_x,F_y,F_z\}$ is a vector whose three components are the expectation
values of the three spin-operators over the multicomponent wavefunction, and 
is called the spin-expectation value \cite{ueda}. Also,  
\begin{align}&
F_{\pm}\equiv  F_x \pm i F_y=
\sqrt 2[\psi_{\pm 1}^*(\mathbf r)\psi_0(\mathbf r)
+\psi_0^*(\mathbf r)\psi_{\mp 1}(\mathbf r)]\label{fpmspin1}, \\
&  F_z= n_{+1}(\mathbf r)-n_{-1}(\mathbf r)\label{fzspin1},
\quad
{\cal H}= -\frac{\nabla^2}{2} +{V}({\mathbf r})+c_0 n,\\
&c_0 = \frac{2N \sqrt{2\pi \gamma}(a_0+2 a_2)}{3 l_0},~
c_1 = \frac{2N \sqrt{2\pi \gamma}({a_2-a_0})}{3 l_0},\label{nonlin}\\
&\nabla^2=\frac{\partial ^2}{\partial x^2} + \frac{\partial ^2}{\partial y^2},~
V(\mathbf r)=\frac{x^2+\beta^2y^2}{2},~
{\mathbf r }\equiv\{x,y\}, \label{nabla_q2d}
\end{align}
where $n_j=|\psi_j(\mathbf r)|^2$ with $j=\pm 1, 0$ are the component densities, 
$n=\sum_{j}n_j$ is the total density, $\mu_j$ with $j=\pm 1, 0$
are the respective chemical potentials, $a_0$ and $a_2$ are the $s$-wave scattering lengths in the 
total spin 0 and 2 channels, respectively, and asterisk denotes complex conjugate.
The normalization condition condition satisfied by the component wavefunctions 
$\psi_j$ is $\int \sum_j n_j d{\bf r} =1$.
Here $l_0=\sqrt{\hbar/(M\omega_x)}$ is the oscillator length along $x$ axis, 
$\beta = \omega_y/\omega_x$ and $\gamma = \omega_z/\omega_x$ are anisotropy parameters. 
Here the units of length, density and chemical potential are $l_0$, $l_0^{-2}$,
and $\hbar\omega_x$, respectively.

 In the same manner, five coupled GP equations, for different spin components
$m_f=\pm 2,\pm 1,0$ of a spin-2 BEC, in dimensionless form are \cite{Koashi,ueda}
 \begin{align} 
 \mu_{\pm 2}& \psi_{\pm 2}(\mathbf r) =
{\cal H}\psi_{\pm 2}(\mathbf r) 
+({{c}_2}/{\sqrt{5}}){\Theta}\psi_{\mp 2}^*(\mathbf r)\nonumber\\
&+{c}_1\big[{F}_{\mp} \psi_{\pm 1}(\mathbf r)\pm 2{F}_{{z}}\psi_{\pm 2}(\mathbf r)\big] 
\label{gp_s1},
\end{align}
\begin{align}
 \mu_{\pm 1}& \psi_{\pm 1}(\mathbf r) = 
{\cal H} \psi_{\pm 1} (\mathbf r) 
-({{c}_2}/{\sqrt{5}}){\Theta}\psi_{\mp 1}^*(\mathbf r)\nonumber\\
 &+{c}_1\big[\sqrt{3/2} {F}_{\mp}\psi_0(\mathbf r)+{F}_{\pm}\psi_{\pm 2} (\mathbf r)
\pm {F}_{{z}}\psi_{\pm 1}(\mathbf r)\big] ,
 \label{gp_s2}
 \\
 \mu_{0}& \psi_{0}(\mathbf r) =
{\cal H}\psi_0(\mathbf r) 
+({{c}_2}/{\sqrt{5}}){\Theta}\psi_{0}^*(\mathbf r)\nonumber\\
  &+c_1\sqrt{3/2}\big[{F}_{-}
 \psi_{-1}(\mathbf r)+{F}_{+}\psi_{+1}(\mathbf r)\big]
  ,\label{gp_s3}
\end{align}
where
\begin{align}
 {F}_{+} =&  {F}_{-}^*= 2(\psi_{+2}^*\psi_{+1}+\psi_{-1}^*\psi_{-2}) \nonumber \\
&+\sqrt{6}(\psi_{+1}^*\psi_0 +\psi_0^*\psi_{-1}),  \\
{F}_{{z}} =& 2(n_{+2}-n_{-2}) + n_{+1}-n_{-1},  \\
{\Theta} =& \frac{2\psi_{+2}\psi_{-2}-2\psi_{+1}\psi_{-1}+\psi_0^2}{\sqrt{5}}.\label{theta}
\end{align}    
Here \begin{align}
c_0 &= 2 N \sqrt{2\pi\gamma}\frac{4a_2+3a_4}{7l_0}, \label{nonlin21}\\
c_1 &=2 N\sqrt{2\pi\gamma}\frac{a_4-a_2}{7l_0},\label{nonlin22}\\
c_2 &= 2 N\sqrt{2\pi\gamma}\frac{7a_0-10a_2+3a_4}{7l_0},\label{nonlin23}\end{align}
 $a_0,a_2$, and   $a_4$ are the $s$-wave scattering lengths in the 
total spin 0, 2 and 4 channels, respectively, and $\mu_j$ with $j=\pm 2,\pm 1, 0$
are the respective chemical potentials. All other variables have 
the same definitions as in the spin-1 case.

In the absence of magnetic field, depending on the values of the interaction parameters $c_j$, the ground state phase of the
spinor BEC is either ferromagnetic or antiferromagnetic (polar) for spin-1 BEC \cite{Ohmi,Ho} and is
ferromagnetic or antiferromagnetic  or cyclic for spin-2 BEC \cite{Ciobanu,ueda}.
%acquires 
%distinct properties and is classified as ferromagnetic, antiferromagnetic or polar, cyclic, etc \cite{Ho,Ciobanu,ueda}.  
%For a spin-1 BEC, $c_1 <0$ corresponds to ferromagnetic phase and $c_1>0$ to polar phase \cite{Ho}. 
%For a spin-2 BEC,   $c_1<0$ and $c_2>20c_1$ correspond to ferromagnetic phase 
%and $c_2<0$,  $c_2<20c_1$ correspond to polar phase, and $c_1>0$ and $c_2>0$ correspond to cyclic phase \cite{Ciobanu,ueda}. 

\section{Fractional-charge vortex dipoles}
\label{3}
The superfluid velocity ${\bf v_s} = \hbar \nabla \theta/M$ for a spinor BEC 
with zero spin expectation value (${\bf F} = 0$), where $\theta$ is the gauge phase of the wave function. 
Hence, antiferromagnetic phase of spin-1, spin-2 and cyclic phase of spin-2 cyclic BECs, each of which 
have ${\bf F} = 0$, are irrotational ($\nabla \times \bf v_s = 0$), and the circulation of the 
velocity field is quantized \cite{ueda}, i.e. $\oint {\bf v_s}.d{\bf l} = n h/M $, 
where $n$ is an integer or a rational fraction. %On the other hand, for spin-1 
%and spin-2 ferromagnetic BECs  the circulation is not quantized \cite{ueda}.
The vortices with %circulation of the velocity field equal to fractional multiple
%of $ h/M $ 
with $n$ equal to a rational fraction are known as fractional-charge vortices, which emerge due to 
the fact that a spinor BEC has SO(3) rotational symmetry along with U(1) global 
gauge symmetry of a scalar BEC \cite{sakurai,ueda}. %Using U(1) global gauge symmetry and SO(3) spin rotational
In the present manuscript, we consider the rotation only around $z$ axis as
the condensate lies in $x$-$y$ plane with no dynamics along $z$ axis, in which case the antiferromagnetic and cyclic phases
of the spinor BEC with non-zero (longitudinal) magnetization 
($ {\cal M} = \int F_z({\bf r}) d\mathbf r\ne 0$) are also irrotational \cite{ueda}, and the concept of fractional 
charge can be extended to these phases of spinor condensate with  non-zero $ {\cal M}$.  
 
\subsection{Spin-1 BEC}
There is a single non-degenerate ground state of the spin-1 antiferromagnetic BEC with non-zero 
${\cal M}$ 
in contrast to two degenerate ground states with $ {\cal M} =0$ \cite{tf}. 
The rotation operator around direction $z$ for the spin-1 BEC is 
\cite{ueda}
 \begin{align}
 {\cal D}(\alpha)=\left( {\begin{array}{ccc}
   e^{-i\alpha} & 0&0 \\ 0&1&0 \\ 0 & 0&   e^{i\alpha} \      
\end{array} } \right).
\end{align}
A general normalized state with ${\cal M}\ne 0$ can be obtained by operating 
${\cal D}(\alpha)$ on the state 
\begin{equation}
\chi_0=\left(\sqrt{\frac{1+{\cal M}}{ 2}},0,\sqrt{\frac{1-{\cal M}}{ 2}}\right)^T
\end{equation}
and can be written
as
\begin{eqnarray} 
\chi &=& e^{i\theta}{\cal D}(\alpha)\chi_0\nonumber\\
     &=& \left[e^{-i(\alpha-\theta)}\sqrt{\frac{1+{\cal M}}{2}},0,e^{i(\alpha+\theta)}\sqrt{\frac{1-{\cal M}}{2}}\right]^T,
\end{eqnarray}
where $\theta$ is the overall gauge phase.  
The simplest way to obtain a single-valued spinor with a fractional gauge charge 
($\theta \ne 0$) is to consider $\theta=\alpha= \beta/2$ \cite{fcv}, where $\beta$ is azimuthal angle. 
The wave function, then,  is
\begin{equation}
\chi = \left[\sqrt{\frac{1+{\cal M}}{2}}, 0,e^{i\beta}\sqrt{\frac{1-{\cal M}}{2}}\right]^T.
\end{equation}
This is a 1/2-1/2 vortex, i.e. 1/2 unit each of gauge and spin charge, located at origin. 
In general 1/2-1/2 vortex located at ($x_1,y_1$), which may or may not
coincide with origin, can be described by wave function 
\begin{equation}
\chi=\left[\sqrt{\frac{1+{\cal M}}{2}}, 0,e^{i\tan^{-1}{\frac{y-y_1}{x-x_1}}}\sqrt{\frac{1-{\cal M}}{2}}\right]^T.
\end{equation} 
If in addition to the vortex at ($x_1,y_1$) there is also an antivortex 
at ($x_2,y_2$) in  component $\psi_{-1}$,
the wavefunction can be written as 
\begin{equation}
\chi_{\rm vd} = \left[\sqrt{\frac{1+{\cal M}}{2}}, 0,e^{i(\tan^{-1}{\frac{y-y_1}{x-x_1}}-\tan^{-1}{\frac{y-y_2}{x-x_2}})}
\sqrt{\frac{1-{\cal M}}{2}}\right]^T,
\end{equation}
which is wavefunction for spin-1 condensate with a half-quantum vortex dipole. 

\subsection{Spin-2 BEC}
{\bf Antiferromagnetic}: There is a single non-degenerate ground state of a non-rotating 
spin-2 antiferromagnetic BEC with non-zero magnetization \cite{tf}.
The rotation operator around z direction for spin-2 BEC is \cite{ueda}
 \begin{align}
{\cal D}(\alpha)=\left( {\begin{array}{ccccc}
   e^{-2i\alpha} & 0&0&0&0\\0  &e^{-i\alpha}&0&0&0\\ 0&0&1&0&0\\
0&0&0&e^{i\alpha}&0 \\   0&0&    0 & 0&   e ^{2i\alpha} \      \end{array} } \right).
\end{align}
One can obtain a general normalized wave function by operating 
${\cal D}(\alpha)$ on the representative state
\begin{equation} 
\chi_0=\left[\frac{\sqrt{2+{\cal M}}}{2},0,0,0,\frac{\sqrt{2-{\cal M}}}{2}\right]^T
\end{equation}
and is given as
\begin{equation}
\chi = e^{i\theta}\left[e^{-2i\alpha}\frac{\sqrt{2+{\cal M}}}{2},0,0,0, e^{2i\alpha}\frac{\sqrt{2-{\cal M}}}{2} \right]^T.
\end{equation}
The single-valuedness of the wave function with a fractional-charge vortex can be 
preserved by considering
 $\theta=\beta/2, \alpha=\beta/4$ to get  
\begin{equation}
\chi=\left[\frac{\sqrt{2+{\cal M}}}{2},0,0,0, 
e^{i\beta}\frac{\sqrt{2-{\cal M}}}{2} \right]  ^T,
\end{equation} which is a 1/2-1/4 vortex or a half-quantum vortex.
Again, as in the case spin-1 condensate, replacing 
$\beta = \tan^{-1}{\frac{y-y_1}{x-x_1}}-\tan^{-1}{\frac{y-y_2}{x-x_2}}$, the aforementioned wavefunction 
describes a 1/2,1/4 vortex dipole.

{\bf Cyclic}: There are two degenerate ground states of a non-rotating spin-2 cyclic BEC for any arbitrary
value of ${\cal M}$ \cite{tf}, and only one of them with $\psi_{+1}=\psi_0=\psi_{-2}=0$ leads
to fractional-charge vortex states. %we consider the possibility of generating fractional-charge vortex 
Thus, a general normalized  wave function under rotation can be written as \cite{fcv} 
\begin{equation}
\chi=e^{i\theta}\left[e^{-2i\alpha}\sqrt{\frac{1+{\cal M}}{3}},0,0, e^{i\alpha}\sqrt{\frac{2-{\cal M}}{3}},0 \right]^T \label{opc}.
\end{equation}
There are four simple ways to maintain a single-valued spinor function to 
generate a fractional-charge vortex \cite{cyclic}:
\begin{align}
\theta&=\beta/3,\alpha=2\beta/3,\label{ta1}\\
\theta&=-\beta/3,\alpha=\beta/3,\label{ta2}\\
\theta&=-2\beta/3, \alpha=2\beta/3\label{ta3}\\
\theta&=2\beta/3,\alpha=\beta/3\label{ta4}.
\end{align}
Replacing $\beta$ by $\tan^{-1}{\frac{y-y_1}{x-x_1}}-\tan^{-1}{\frac{y-y_2}{x-x_2}}$ in Eqs. (\ref{ta1})-(\ref{ta4})
and then substituting in Eq. (\ref{opc}), one can obtain the wavefunctions describing 
1/3- and 2/3-quantum vortex dipoles in cyclic spin-2 condensate.

In all the cases discussed above of fractional-charge vortex dipoles, 
the spinor BEC has all but two of the component wavefunctions zero.
Thus, the GP equations for the spinor BEC simplify
to that for a two-component BEC \cite{fcv} and can be written as \cite{fcv}
\begin{align}\label{bi}
&i\frac{\partial}{\partial t}\phi_j =\left[-\frac{\nabla^2}{2}+\frac{x^2+y^2}{2}
+ g_j\phi_j^2 + g_{12}\phi_{3-j}^2\right]\phi_j,\\
&\nabla^2 = \frac{\partial}{\partial x^2} +\frac{\partial}{\partial y^2}, 
\end{align} where $j=1,2$, $g_j$ and $g_{12}$ are the intra-component and inter-component interaction parameters, respectively.
For antiferromagnetic phase of spin-1 and spin-2 BEC  with a half-quantum vortex dipole
and coupling the two non-zero components, these interaction parameters are 
\begin{align} 
g_1 &= g_2 = c_0 + c_1,~g_{12}=c_0-c_1.\label{nl4}\\
g_1 &=g_2=c_0+4c_1,~g_{12}=c_0-4c_1+2c_2/5,\label{nl2}
\end{align}
respectively \cite{fcv}. 
Similarly, for a spin-2 cyclic BEC with a fractional-charge vortex dipole and 
coupling $\psi_{+2}=\phi_1$ and $\psi_{-1}=\phi_2$, the interaction parameters are \cite{fcv}
\begin{equation}
 g_1 = c_0 + 4c_1,~g_2 = c_0 + c_1,~ g_{12} = c_0 -2c_1. \label{nl1}
\end{equation}
Eqs. (\ref{nl4})-(\ref{nl1}) are the definitions of intra- and 
inter-species interaction parameters of a two-component BEC which 
can be treated as equivalent to a spinor BEC hosting a fractional-charge
vortex dipole.

\section{Critical speeds for the binary system}
\label{sec-IV}
The critical speed for vortex shedding $v_c$ in compressible fluid by a penetrable obstacle is 
a fraction of the local speed of sound \cite{Kwon,Huepe}. Hence, in order to understand
the emergence of two different critical speeds for vortex dipole shedding in antiferromagnetic
and cyclic phases of spinor condensate, we calculate the excitation spectrum of the uniform
binary system, introduced in the previous section, representing the spinor BEC. It must be
stated that there are a few spin-wave excitations which are not captured by the analysis 
based on the binary system, but they do not play any role in the vortex shedding dynamics 
considered in this paper and are decoupled from the modes captured by binary system 
approach \cite{Watabe}. If $\delta \phi_j$ are the changes in the wavefunctions from their
respective unperturbed values, one can derive the Bogoliubov-de Gennes (BdG) equations for 
the uniform binary condensate ($V = 0$) by linearizing Eqs. (\ref{bi}) around the unperturbed 
wavefunctions, i.e., substituting $\phi_j = \bar{\phi}_j + \delta \phi_j$ and retaining
terms only upto first order in $\delta \phi_j$, we get
\begin{eqnarray}
i\frac{\partial \delta\phi_j}{\partial t} &=& -\frac{\nabla^2}{2}\delta\phi_j
+g_j\left( 2|\bar{\phi}_j|^2\delta \phi_j+\bar{\phi}_j^2\delta \phi_j^*\right) +g_{12}\label{le1}\\&&
 \times (|\bar{\phi}_{3-j}|^2\delta\phi_j+\bar{\phi}_j\bar{\phi}_{3-j}\delta\phi_{3-j}^*+
\bar{\phi}_j\bar{\phi}_{3-j}^*\delta\phi_{3-j})\nonumber
\end{eqnarray} 
where $\bar{\phi}_j$ are to be understood as equilibrium values of wavefunctions.
Now, we substitute $\bar{\phi}_j=\sqrt{n_j}e^{-i\mu_j t}$ and 
$\delta \phi_j = e^{-i\mu_j t}[u_j(\mathbf r)e^{-i\omega t}-v_j^*(\mathbf r)e^{i\omega t}]$ with
$u_j(\mathbf r)$ and $v_j(\mathbf r)$ denoting the quasi-particle amplitudes, Eqs. (\ref{le1}) then become
\begin{widetext}
\begin{eqnarray}
(\mu_j+\omega)u_je^{-i\omega t}-(\mu_j-\omega)v_j^*e^{i\omega t}=
-\frac{\nabla^2}{2}\left( u_j e^{-i\omega t} - v_j^*e^{i\omega t}\right)
+g_j[ 2n_j( u_j e^{-i\omega t} -v_j^*e^{i\omega t})+n_j(u_j^*e^{i\omega t}-v_je^{-i\omega t})]\nonumber\\
 +g_{12}[n_{3-j}(u_je^{-i\omega t}-v_j^*e^{i\omega t})+\sqrt{n_1n_2}(u_{3-j}^*e^{i\omega t}-v_{3-j}e^{-i\omega t}
+u_{3-j}e^{-i\omega t}-v_{3-j}^*e^{i\omega t})] \label{le2},
\end{eqnarray}
\end{widetext}
Finally, we equate the coefficients of $e^{\mp i\omega t}$ on both sides of Eqs. (\ref{le2}).
The coupled BdG equations thus obtained are
\begin{eqnarray}
A(u_1,v_1,u_2,v_2)^T = \omega(u_1,v_1,u_2,v_2)^T,
\end{eqnarray}
where $A$ is a $4\times4$ matrix whose elements for a uniform system are given as \cite{BdG}
\begin{eqnarray}
A_{11} &=& -\frac{\nabla^2}{2}+2 g_1 n_1+g_{12} n_2-\mu_1,
A_{12} =-g_1 n_1\nonumber\\
A_{13} &=& g_{12}\sqrt{n_1n_2},
A_{14} = -g_{12}\sqrt{n_1n_2}\nonumber\\
A_{21} &=& g_1n_1,
A_{22} = -(-\frac{\nabla^2}{2}+2g_1 n_1+g_{12} n_2)+\mu_1\nonumber\\
A_{23} &=& g_{12}\sqrt{n_1n_2},
A_{24} = -g_{12}\sqrt{n_1n_2}\nonumber\\
A_{31} &=&g_{12}\sqrt{n_1n_2},
A_{32} =-g_{12}\sqrt{n_1n_2}\nonumber\\
A_{33} &=&-\frac{\nabla^2}{2}+2g_2 n_2+g_{12} n_1-\mu_2,
A_{34} =-g_2n_2\nonumber\\
A_{41} &=&g_{12}\sqrt{n_1n_2},
A_{42} =-g_{12}\sqrt{n_1n_2}\nonumber\\
A_{43} &=&g_2n_2,
A_{44} =-(-\frac{\nabla^2}{2}+2g_2 n_2+g_{12} n_1)+\mu_2.\nonumber
\end{eqnarray}
For a uniform system with $\mu_1 = g_1 n_1+g_{12} n_2$ and $\mu_2 = g_2 n_2+g_{12}n_1$, 
seeking solutions of form $u(\mathbf r) = u_q e^{i{\bf q} .{\bf r}}/\sqrt{{\cal V}}$ and
$v(\mathbf r) = v_q e^{i{\bf q}.{\bf r}}/\sqrt{{\cal V}}$ where ${\cal V}$ is the volume (area)
in three (two)-dimensional space, the excitation spectrum for the consistent solutions
of BdG equations is
\begin{widetext}
\begin{eqnarray}
\omega_{\pm }&=&\frac{q}{2} \sqrt{ 2 g_1 n_1+ 2 g_2 n_2 \pm
2 \sqrt{ \left(g_1^2 n_1^2-2 g_1g_2 n_1 n_2+4 g_{12}^2 n_1n_2+g_2^2n_2^2\right)}+ q^2}\label{es1}.
\end{eqnarray}
\end{widetext}
For spin-1 antiferromagnetic condensate, using $n_1 = (1+{\cal M})n/2, n_2 = (1-{\cal M})n/2$ where $n$ is the total
density of the uniform system, and Eq. (\ref{nl4}) in Eq. (\ref{es1}), we get
\begin{widetext}
\begin{equation}
\omega_{\pm } = \frac{q \sqrt{2 c_0n+2 c_1 n\pm2 n \sqrt{c_0^2-2 c_0 c_1+c_1^2+4 c_0 c_1 {\cal M}^2}+q^2}}{2}\label{es2}.
\end{equation} 
\end{widetext}
It has been shown in Ref. \cite{ueda} that these two excitations are coupled and correspond to 
coupled phonon and magnon excitations, with two critical speeds
\begin{eqnarray}
c_{\pm}&=&\lim_{q\rightarrow 0}\frac{\omega_{\pm}(q)}{q},\\ 
       &=& \frac{\sqrt{2 c_0n+2 c_1 n\pm2 n \sqrt{c_0^2-2 c_0 c_1+c_1^2+4 c_0 c_1 {\cal M}^2}}}{2}\label{cpm_spin1_polar}.
\end{eqnarray}
Depending upon whether ${\cal M}$ is zero or not zero, the two critical speeds can lead
to single, $v_c$, or two different, $v_c^l$ and $v_c^h$, critical speeds for vortex dipole shedding by a moving obstacle potential.
If ${\cal M} = 0$, the two modes become decoupled with excitation frequency given by
\begin{eqnarray}
\omega_{+} &= \frac{q \sqrt{4 c_0 n+q^2}}{2},\\
\omega_{-} &= \frac{q \sqrt{4 c_1 n+q^2}}{2},
\end{eqnarray}
with two critical speeds
\begin{eqnarray}
c_{+} &= \sqrt{c_0 n}\\
c_{-} &= \sqrt{c_1 n}.
\end{eqnarray}
The decoupling of the critical speeds for phonon $c_{+}$ and magnon $c_{-}$ excitations at ${\cal M} = 0$
results in a single critical speed $v_c$ for the vortex-antivortex nucleation by a moving obstacle
and corresponds to critical speed for phonon excitation which can lead to density
fluctuations. As mentioned earlier, for a penetrable obstacle potential $v_c$ should be the fraction
of $c_{+}$. In this case, the critical speed for spin excitations $c_{-}$ is not associated
with the vortex shedding phenomenon. On the other hand, coupling of the two modes for ${\cal M} \ne 0$ results
in two critical speeds, $v_c^l$ and $v_c^h$ with $v_c^l<v_c^h$, for the vortex-antivortex nucleation by a moving obstacle. This
coupling of the modes can be used to create half-quantum vortex dipole in spin-1 antiferromagnetic
condensate. Again $v_c^l$ and $v_c^h$ are fractions of $c_{-}$ and $c_{+}$, respectively. 
Similarly, the two critical speeds for antiferromagnetic and cyclic spin-2 condensate
are, respectively,
\begin{widetext}
\begin{eqnarray}
c_{\pm} &=& \frac{\sqrt{10 c_0n+40 c_1 n\pm2  n \sqrt{25 c_0^2+400 c_1^2-20 c_1 c_2
\left(4-{\cal M}^2\right)+c_2^2 \left(4-{\cal M}^2\right)+5 c_0 \left(c_2 \left(4-{\cal M}^2\right)
+20 c_1 \left({\cal M}^2-2\right)\right)}}}{2 \sqrt{5}}\\
c_{\pm} &=& \frac{\sqrt{2 c_0 n+4 c_1  n+2 c_1  n {\cal M}\pm2  n \sqrt{c_0^2+c_1^2 (2+{\cal M})^2+
2 c_0c_1 \left(2 {\cal M}^2-{\cal M}-2\right)}}}{2}.
\end{eqnarray}
\end{widetext}
As in case of antiferromagnetic spin-1 condensate, the phonon and spin modes gets decoupled 
for ${\cal M} = 0$, resulting in single critical speed for vortex-antivortex pair
creation by a moving obstacle.

\section{Numerical results}
\label{V}
We use the time-splitting Fourier pseudo-spectral method to solve the full GP equations
for spin-1 and spin-2 condensate \cite{num}. The method automatically implements the periodic
boundary conditions used to simulate the uniform system. Alternatively, one can also implement
the time-splitting Crank-Nicolson method with explicit periodic boundary conditions to simulate
uniform system and box boundary conditions to simulate trapped condensates \cite{cn}. 
We use the imaginary time propagation to obtain
the ground state solution and real time propagation to study the dynamics of fractional-charge
vortex dipole generation. The space and time steps employed in our study are $0.1$ and 
$0.0005$, respectively. 
We consider spin-1 and spin-2 condensates each of $^{23}$Na atoms for numerical 
simulations. The two scattering lengths of spin-1 $^{23}$Na condensate are 
$a_0=47.36a_B$ and $a_2=52.98a_B$ \cite{ueda}, and the three scattering lengths of spin-2 
condensate of $^{23}$Na atoms are $a_0 = 34.9a_B$, $a_2 = 45.8a_B$, and 
$a_4 = 64.5a_B$ \cite{Ciobanu}, where $a_B$ is Bohr radius. The ground state phases of both spin-1 and
spin-2 $^{23}$Na condensate with these scattering length values are antiferromagnetic.
In order to access the cyclic phase of spin-2 condensate, we reduce $a_2$ to 
half of its actual value which can be achieved experimentally by exploiting 
Feshbach resonances \cite{Inouye}. We consider both the uniform and trapped spinor condensates in
our simulations. 

\begin{figure}[h]
\begin{center}
\includegraphics[trim = 30mm 5mm 30mm 30mm, clip,width=1\linewidth,clip]{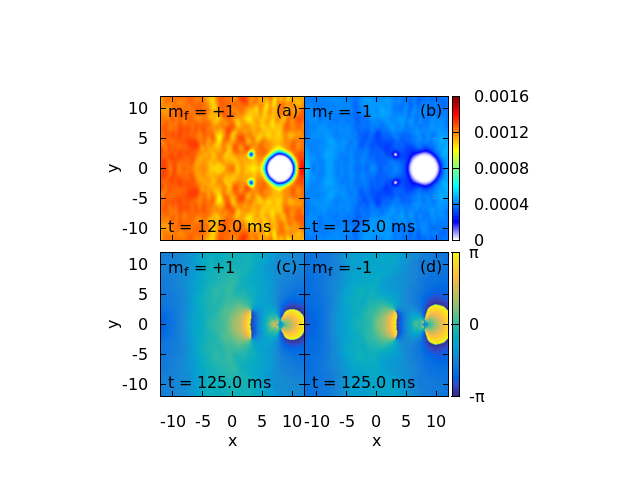}
\caption{(Color online) Nucleation of a gauge vortex dipole in uniform antiferromagnetic
spin-1 condensate of $^{23}$Na with $c_0 = 6466.5$, $c_1 = 237.0$, and ${\cal M} = 0.5$. 
(a)-(b) and (c)-(d) show the density and phase plots, respectively. All the variables in this and the other figures
in this paper are in dimensionless units.}
\label{fig1}
\end{center}
\end{figure}

{\em Uniform system:} %Here with $50000$ atoms distributed over a square of size $25.5\times 25.5$, 
We first consider spin-1 antiferromagnetic condensate over a square of size $25.5\times 25.5$ 
with $c_0 = 6466.5$ and $c_1 = 237.0$. The values of nonlinearities correspond to choosing $N=50000$, $l_0=4.69\mu$m (unit
of length for numerical simulations), $\gamma = 20$,
$a_0=47.36a_B$, and $a_2=52.98a_B$ in Eqs. (\ref{nonlin}). We move the Gaussian obstacle potential 
$V_{\rm obs} = V_0e^{-2[(x-x_0(t))^2+y^2]/w^2}$, initially located at $x_0(0) = -8.0$,
with a constant speed $v$ such that $x_0(t) = x_0(0) +vt$, where $V_0$ and $w$ are the strength and width of the 
obstacle potential. In spin-1 condensate, we consider $V_0=100$ (in dimensionless units), 
$w_0=10\mu$m. The obstacle is moved with a speed of $0.6$ mms$^{-1}$ across the condensate
with ${\cal M}=0.5$. We find that a vortex dipole is created in both the components as is
shown in Fig. \ref{fig1}, which corresponds to a gauge vortex dipole in scalar condensates.
One unit of length in this and other figures in this paper is equal to $4.69\mu$m. 
This indicates (almost) same critical speed for vortex-antivortex pair creation in this case.
This is due to the fact that in this case $g_{12}=c_0-c_1\approx g_1=g_2$ which results in a single 
non-zero critical speed $c_{+}$ from Eq. (\ref{es1}). The critical speed for vortex dipole shedding 
obtained from the 
numerical simulations is $v_c\approx 0.59$mms$^{-1}$ which is a fraction (0.32 times) of the 
speed of the "sound" $c_{+}\simeq1.86$mms$^{-1}$ obtained from Eq. (\ref{cpm_spin1_polar}). 
This is in good agreement with the estimation for the critical speed in compressible 
superfluids \cite{Kwon,Huepe,Hakim}. 

\begin{figure}[h]
\begin{center}
\includegraphics[trim = 30mm 5mm 30mm 30mm, clip,width=1\linewidth,clip]{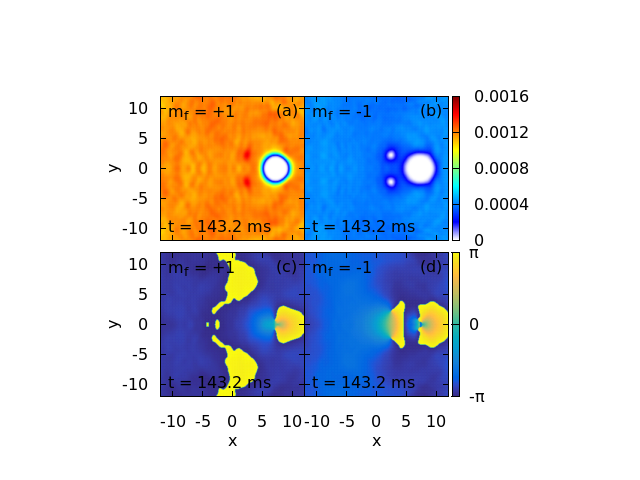}
\caption{(Color online) Nucleation of a 1/2,1/2 fractional vortex dipole in uniform antiferromagnetic
spin-1 condensate of $^{23}$Na with $c_0 = 8701.0$, $c_1 = 1354.3$, 
and ${\cal M} = 0.5$.(a)-(b) and (c)-(d) show the density and phase plots, respectively.}
\label{fig2}
\end{center}
\end{figure}

In order to confirm the coupling
between the phonon and magnon modes, we consider antiferromagnetic spin-1 condensate with $c_0=8701.0$
and $c_1 = 1354.3$ over a box of size $25.5\times 25.5$. %change $a_2$ to $1.5$ times of its actual value.
The values of nonlinearities correspond to choosing $N=50000$, $l_0=4.69\mu$m, $\gamma = 20$,
$a_0=47.36a_B$, and $a_2=79.47a_B$ in Eqs. (\ref{nonlin}). 
In this case, moving the same obstacle potential at
a speed of $0.5$ mms$^{-1}$ across the condensate with ${\cal M} = 0.5$ leads to the shedding 
of vortex dipole in $m_f=-1$ component which is $1/2,1/2$ vortex dipole as is shown in 
Fig. \ref{fig2}. 
Numerically, we find that there are two critical speeds
$v_c^l\approx0.37$ mms$^{-1}$ and $v_c^h\approx0.63$ mms$^{-1}$ for vortex dipole shedding in this case, 
and $0.5$ mms$^{-1}$ is greater than the smaller of the two. When moved with a speed greater 
than $0.63$ mms$^{-1}$, $1/2,1/2$ vortex dipoles with singularities in $m_f = 1$ component 
are also created. The two critical speeds are fractions
of the two speeds of "sound", $\simeq0.715$ mms$^{-1}$  and $\simeq2.185$ mms$^{-1}$ obtained 
from Eq. (\ref{cpm_spin1_polar}). The size of the vortex in a condensate is of the order of 
coherence length, which for a binary condensate is
$
\xi_j = 1/\sqrt{2(g_jn_j+g_{12}n_{3-j})}=1/\sqrt{2\mu_j},
$ where $n_j$ and $n_{3-j}$ are the uniform component densities.
Using this relation, the size of the vortex cores is $0.2$ in this case.
 
\begin{figure}[h]
\begin{center}
\includegraphics[trim = 30mm 5mm 30mm 30mm, clip,width=1\linewidth,clip]{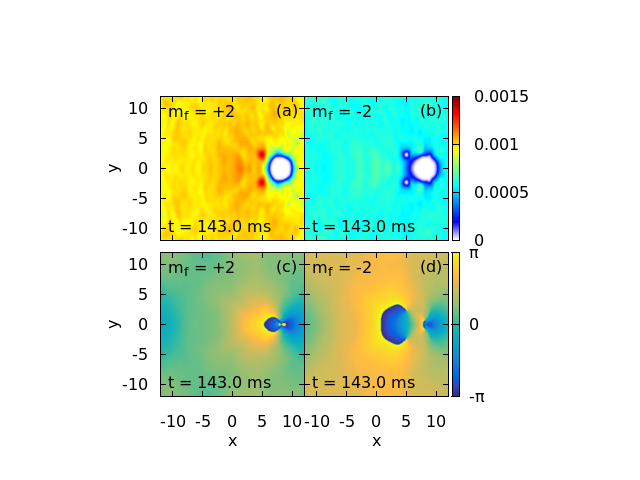}
\caption{(Color online) Nucleation of a 1/2,1/4 fractional vortex dipole in uniform antiferromagnetic
spinor condensate of $^{23}$Na with $c_0 = 6809.1$, $c_1 = 338.0$, 
$c_2 = -365.1$, and ${\cal M} = 0.5$. (a)-(b) and (c)-(d) show the density
and phase plots, respectively.}
\label{fig3}
\end{center}
\end{figure}

\begin{figure}[h]
\begin{center}
\includegraphics[trim = 30mm 5mm 30mm 30mm, clip,width=1\linewidth,clip]{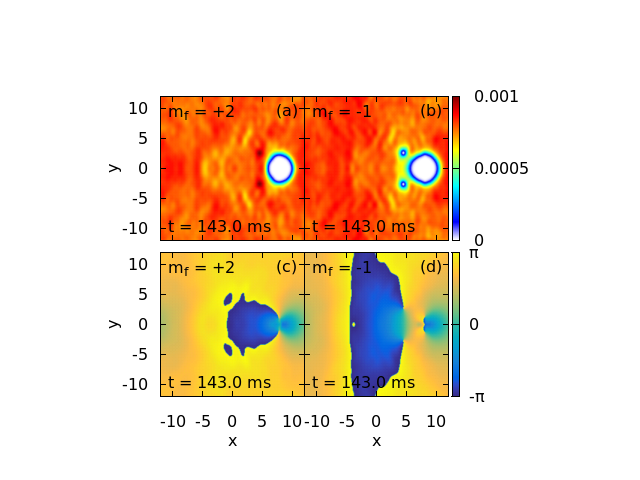}
\caption{(Color online) Nucleation of a 2/3,1/3 fractional vortex dipole in uniform cyclic 
spinor condensate of $^{23}$Na with $c_0 = 5153.4$, $c_1 = 751.9$, 
$c_2 = 3774.2$, and ${\cal M} = 0.5$. (a)-(b) and (c)-(d) show the density
and phase plots, respectively.}
\label{fig4}
\end{center}
\end{figure}

In the case of antiferromagnetic spin-2 condensate, we consider 
$c_0 = 6809.1$, $c_1 = 338.0$, $c_2=-365.1$ over a square of size $25.5\times25.5$. 
The values of nonlinearities correspond to choosing 
$N=50000$, $l_0=4.69\mu$m, $\gamma = 20$, $a_0=34.9a_B$, $a_2=45.8a_B$, $a_4=64.5$ in Eqs. (\ref{nonlin21}-\ref{nonlin23}).
Similarly, for the cyclic spin-2 condensate, we consider  $c_0 = 5153.4$, $c_1 = 751.9$, $c_2=3774.2$, which
correspond to choosing $N=50000$, $l_0=4.69\mu$m, $\gamma = 20$, $a_0=34.9a_B$, $a_2=27.9a_B$, 
$a_4=64.5$ in Eqs. (\ref{nonlin21}-\ref{nonlin23}).
For both the antiferromagnetic and cyclic phases, we consider
$V_0 =50$  and $w_0 = 10\mu$m. Here moving the obstacle potential with
a speed of $0.525$ mms$^{-1}$ across the condensate with ${\cal M} = 0.5$ leads to the 
creation of $1/2,1/4$ and $2/3,1/3$ vortex dipoles in the antiferromagnetic and cyclic
phases respectively; these are shown in Fig. \ref{fig3} and Fig. \ref{fig4}, respectively.
The coherence lengths as a measure of the size of the vortex cores are $0.23$ and $0.26$ for
the antiferromagnetic and cyclic phases, respectively. 

\begin{figure}[t]
\begin{center}
\includegraphics[trim = 10mm 10mm 10mm 30mm, clip,width=0.95\linewidth,clip]{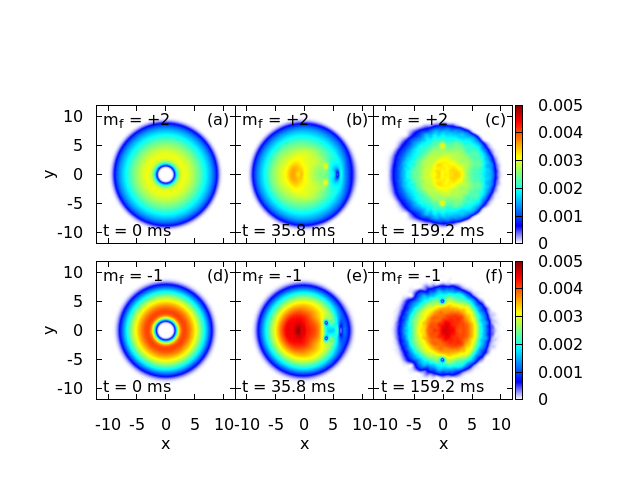}
\caption{(Color online) Creation of 2/3,1/3 vortex dipole in a cyclic spinor condensate
of $50000$ $^{23}$Na atoms with $a_0 = 34.9a_B$, $a_2 = 27.9a_B$,
$a_4 = 64.5a_B$,  and ${\cal M} = 0.5$. 
}
\label{fig5}
\end{center}
\end{figure}

{\em Trapped system:} To illustrate the creation of fractional-charge vortex dipoles in trapped systems, 
we consider $5\times10^4$ spin-2 $^{23}$Na atoms trapped in quasi-2D trap with 
$\omega_x=\omega_y=2\pi\times20$, $\omega_z=2\pi\times400$ along with a Gaussian obstacle
potential $V_{\rm obs}$ initially located at $x_0(0) = 0$. The oscillator length $l_0 = 4.69\mu$m
with these set of parameters. Here while moving the obstacle 
potential along $x$ axis with a constant speed, its strength is continuously decreased 
at a constant rate in order to make $V_{\rm obs}$ vanish at some desired point on the 
$x$ axis. We consider $V_0 = 100\hbar\omega_x$ as the initial strength of the obstacle potential.
The obstacle is moved along $x$ axis with a speed of $0.6$ mms$^{-1}$. The obstacle traverses
a distance of $5a_{\rm osc}$ before it vanishes. 
The generation of a single 2/3,1/3 fractional vortex
dipole in the cyclic phase with ${\cal M} = 0.5$ is shown in Fig. \ref{fig5}.
Similarly, we were able to generate half-quantum vortices in the antiferromagnetic phases
of spin-1 and spin-2 condensates of $50000$ $^{23}$Na atoms each with ${\cal M} = 0.5$ and
scattering length values same as in Figs. \ref{fig2} and  \ref{fig3}.

\begin{figure}[t]
\begin{center}
\includegraphics[trim = 10mm 10mm 10mm 30mm, clip,width=0.95\linewidth,clip]{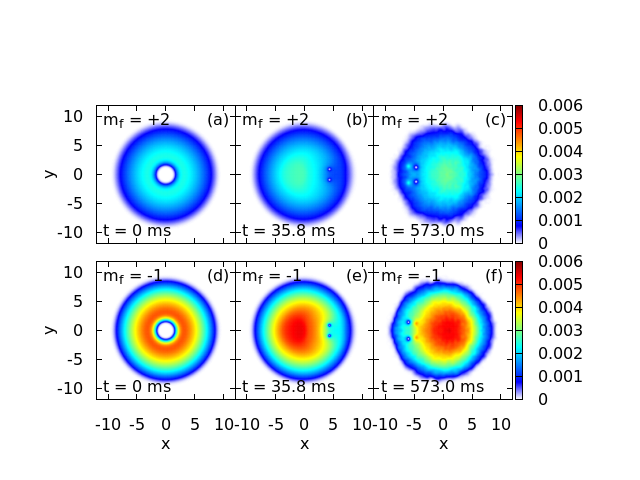}
\caption{(Color online) Creation of 1/3,1/3 (dipole in component $m_f=+2$)
and 2/3,1/3 (dipole in component $m_f=-1$) vortex dipoles in a cyclic spin-2 condensate
of $50000$ $^{23}$Na atoms with $a_0 = 34.9a_B$, $a_2 = 27.9a_B$,
$a_4 = 64.5a_B$,  and ${\cal M} = 0$. (a) and (d), (b) and (e), and (c) and (f) show, respectively, 
the initial density profiles, the density profiles with a gauge vortex dipole in each component, and 
the density profiles with a 1/3- and 2/3-quantum vortex dipoles.}
\label{fig6}
\end{center}
\end{figure}

\begin{figure}[h]
\begin{center}
\includegraphics[trim = 10mm 10mm 10mm 30mm, clip,width=0.95\linewidth,clip]{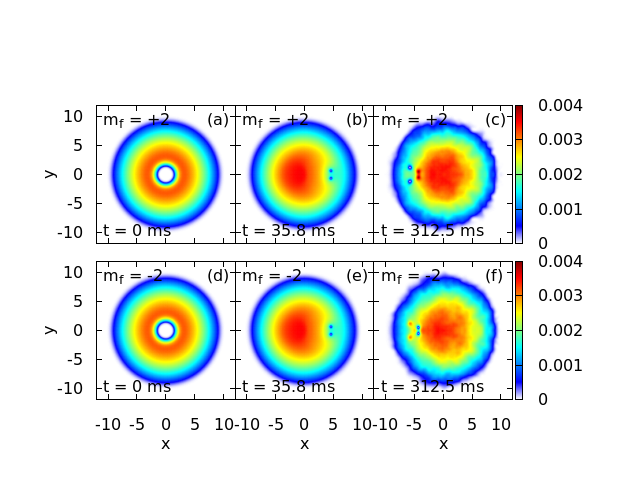}
\caption{(Color online) Nucleation of 1/2,1/4 vortex dipoles, two pairs, in an antiferromagnetic spin-2
condensate of $50000$ $^{23}$Na atoms with $a_0 = 34.9a_B$, $a_2 = 45.8a_B$,
$a_4 = 64.5a_B$ and ${\cal M} = 0.005$. (a) and (d), (b) and (e), and (c) and (f) show, respectively, 
the initial density profiles, the density profiles with a gauge vortex dipole in each component, and 
the density profiles with a 1/2,1/4 vortex dipole in each component.}
\label{fig7}
\end{center} 
\end{figure}

In both the antiferromagnetic and cyclic phases, reducing the ${\cal M}$ to zero leads to the
generation of gauge vortex dipoles as is shown in Figs. \ref{fig6} (b) and (e) for the cyclic phase.
This is consistent with fact that the phonon excitation mode gets decoupled from the 
spin excitation mode as discussed in the previous section, which results in a single critical 
speed for vortex-antivortex pair creation in the system. Interestingly, the gauge vortices, shown in
Figs. \ref{fig6} (b) and (e), later on decay into a pair of 1/3 and 2/3 fractional-charge vortices
as is shown in Figs. \ref{fig6} (c) and (f). This dynamical instability
of the gauge vortices is consistent with a recent experiment where the decay of a gauge vortex 
into two half-quantum vortices in an antiferromagnetic spin-1 condensate was observed \cite{Seo},
and is due to the fact that a gauge vortex is energetically unstable to decay into two 
half-quantum vortices in the antiferromagnetic spin-1 condensate \cite{Ji}.
We find that in some cases a small perturbation, like a small non-zero magnetization, 
is needed to initiate the decay of the gauge vortex dipole into a pair of fractional-charge vortices. 
For example, the generation of a pair of $1/2,1/4$ vortex dipoles created in the antiferromagnetic phase 
of the spin-2 $^{23}$Na condensate with ${\cal M} = 0.005$ is shown in Fig. \ref{fig7}. In this case,
we introduced a small difference in the population of the atoms in $m_f=+2$ and $m_f=-2$ sublevels to
initiate the decay of the gauge vortices, shown in Figs. \ref{fig7}(b) and (e), into a 
pair of 1/2,1/4 vortices shown in Figs. \ref{fig7}(c) and (f).
   
\section{Summary and discussion}
\label{VI}
We show that the coupling of the phonon excitation mode and a spin excitation mode in antiferromagnetic
phase of spin-1 and antiferromagnetic and cyclic phases of spin-2 condensates with non-zero magnetization 
results in two critical velocities for vortex dipole shedding in these systems. When an
obstacle potential is moved with a velocity which is in between these two velocities, vortex
dipoles are shed only in one of the components of the spinor condensate. These vortex dipoles
with phase singularities appearing in only one of two non-zero components of the spinor condensate
are vortex dipoles with fractional charge. 
The coupling is absent in antiferromagnetic and cyclic phases with zero magnetization resulting 
in a single critical speed for vortex dipole shedding and is a fraction of the critical speed for
phonon excitation mode. Here, moving the obstacle potential initially generates vortex dipoles in 
both the components, also the phase singularities corresponding to vortex dipole in one component 
coincide with the phase-singularities in the other non-zero component. These are usual gauge vortex 
dipoles investigated earlier in antiferromagnetic spin-1 condensate \cite{Rodrigues}. We find
that a small perturbation like a very small non-zero magnetization can lead to the decay
of the an initially formed gauge vortex dipole into two fractional-charge vortex dipoles.
This can have implications in future experiments on these systems since notwithstanding
the fact that the global minima in energies of antiferromagnetic and cyclic phases of the spinor condensates
occurs for zero magnetization, the condensates prepared in experiments will invariably have
some remnant non-zero magnetization which remains conserved over the life time of the 
condensate 
\cite{ueda}.

\begin{acknowledgments}
 This work is financed by FAPESP (Brazil) under Contract No. 2013/07213-0.
 I thank Prof. S. K. Adhikari for useful discussions. 
\end{acknowledgments}


\begin{thebibliography}{99}
\bibitem{Donnelly}
 R. J. Donnelly, {\em Quantized Vortices in Helium II} (Cambridge
University Press, Cambridge, UK, 1991).
\bibitem{ueda}
 Y. Kawaguchi and M. Ueda,
 Phys. Rep. {\bf 520}, 253 (2012).
{\bibitem{Stamper-Kurn}
 D. M. Stamper-Kurn and M. Ueda,
 Rev. Mod. Phys. {\bf 85}, 1191 (2013).}
\bibitem{Masahito}
 M. Ueda,
 Rep. Prog. Phys. {\bf 77}, 122401 (2014).

\bibitem{Onsager}L. Onsager, Nuovo Cimento Suppl. {\bf 6}, 249 (1949);
 R. P. Feynman, Prog. Low Temp. Phys. {\bf 1}, 17 (1955).
\bibitem{Fetter}
 A. L. Fetter,
 Rev. Mod. Phys. {\bf 81}, 647 (2009).

\bibitem{Mizushima-1}
 T. Mizushima, K. Machida, and T. Kita,
 Phys. Rev. Lett. {\bf 89}, 030401 (2002).
\bibitem{Mizushima-2}
 T. Mizushima, K. Machida, and T. Kita,
 Phys. Rev. A {\bf 66}, 053610 (2002).
\bibitem{Mizushima-3}
 T. Mizushima, N. Kobayashi, and K. Machida,
 Phys. Rev. A {\bf 70}, 043613 (2004).
{\bibitem{Leanhardt}
 A. E. Leanhardt, Y. Shin, D. Kielpinski, D. E. Pritchard, and W. Ketterle,
 Phys. Rev. Lett. {\bf 90}, 140403 (2003).}
\bibitem{Ohmi}
 T. Ohmi, K. Machida, J. Phys. Soc. Japan {\bf 67}, 1822 (1998).
\bibitem{Ho}
 T.-L. Ho, Phys. Rev. Lett. {\bf 81}, 742 (1998).
\bibitem{Koashi}
 M. Koashi and M. Ueda,
 Phys. Rev. Lett. {\bf 84}, 1066 (2000);
  M. Ueda and M. Koashi, Phys. Rev. A {\bf 65}, 063602 (2002).
\bibitem{Ciobanu}
C. V. Ciobanu, S.-K. Yip, and T.-L. Ho,
Phys. Rev. A {\bf 61,} 033607 (2000).
\bibitem{Zhou}
 F. Zhou,
 Int. J. Mod. Phys. B {\bf 17}, 2643 (2003);
 U. Leonhardt and G.E. Volovik,
 JETP Lett. {\bf 72}, 46 (2000);
 A.-C. Ji, W. M. Liu, J. L. Song, and F. Zhou,
 Phys. Rev. Lett. {\bf 101,} 010402 (2008).
\bibitem{Seo}
 S. W. Seo, S. Kang, W. J. Kwon, and Y.-il Shin,
 Phys. Rev. Lett. {\bf 115}, 015301 (2015).
\bibitem{Isoshima}
 T. Isoshima, K. Machida, and T. Ohmi,
 J. Phys. Soc.  Japan {\bf 70}, 1604 (2001);
 T. Isoshima and K. Machida,
 Phys. Rev. A {\bf 66}, 023602 (2002).

\bibitem{Sadler}
 L. E. Sadler, J. M. Higbie, S. R. Leslie, M. Vengalattorem and D. M. Stamper-Kurn,
 Nature (London) {\bf 443}, 312 (2006).
\bibitem{cyclic}J. A. M. Huhtam\"aki, T. P. Simula,  M. Kobayashi,  and K. Machida,
Phys. Rev. A {\bf 80}, 051601(R) (2009).
\bibitem{Tsubota}
 M. Kobayashi and M. Tsubota,
 Phys. Rev. Lett. {\bf 94}, 065302 (2002);
 T.-L. Horng, C.-H. Hsueh, and S.-C. Gou
 Phys. Rev. A {\bf 77}, 063625 (2008);
 M. Tsubota, M. Kobayashi, and H. Takeuchi, Phys. Rep. {\bf 522},
191 (2013).
\bibitem{White}
 A. C. White, B. P. Anderson, and V. S. Bagnato, Proc. Natl.
Acad. Sci. USA {\bf 111}, 4719 (2014).
\bibitem{Berezinskii}
 V. L. Berezinskii, Sov. Phys. JETP {\bf 34}, 610 (1972).
\bibitem{Kosterlitz} 
 J. M. Kosterlitz and D. J. Thouless, J. Phys. C {\bf 6}, 1181 (1973);
 J. M. Kosterlitz, {\em ibid.} {\bf 7}, 1046 (1974).
\bibitem{Hadzibabic}
 Z. Hadzibabic, P. Kr\"uger, M. Cheneau, B. Battelier, and J. Dalibard,
 Nature {\bf 441}, 1118 (2006).
\bibitem{Kibble-Zurek}
 T. W. B. Kibble, J. Phys. A {\bf 9}, 1387 (1976);
 W. H. Zurek, Nature (London) {\bf 317}, 505 (1985).
\bibitem{Frisch}
 T. Frisch, Y. Pomeau, and S. Rica, Phys. Rev. Lett. {\bf 69}, 1644
(1992); B. Jackson, J. F. McCann, and C. S. Adams, Phys. Rev. Lett. {\bf 80},
3903 (1998);T. Winiecki, J. F. McCann, and C. S. Adams,
Phys. Rev. Lett. {\bf 82}, 5186 (1999).
\bibitem{Landau}
 L. D. Landau, J. Phys. USSR {\bf 5}, 71 (1941).
\bibitem{Neely}
 S. Inouye, S. Gupta, T. Rosenband, A. P. Chikkatur, A. Gorlitz,
 T. L. Gustavson, A. E. Leanhardt, D. E. Pritchard, and
 W. Ketterle, Phys. Rev. Lett. {\bf 87}, 080402 (2001);
 T. W. Neely, E. C. Samson, A. S. Bradley, M. J. Davis, and
 B. P. Anderson, Phys. Rev. Lett. {\bf 104}, 160401 (2010);
 W. J. Kwon, G. Moon, J. Choi, S. W. Seo, and Y. Shin, Phys.
 Rev. A 90, 063627 (2014).
\bibitem{Kwon}
 M. Crescimanno, C. G. Koay, R. Peterson, and R. Walsworth,
 Phys. Rev. A {\bf 62}, 063612 (2000);
 W. J. Kwon, G. Moon, S. W. Seo, and Y. Shin,
 Phys. Rev A {\bf 91}, 053615 (2015).

{ \bibitem{Salasnich}
 L. Salasnich, A. Parola, and L. Reatto, 
 Phys. Rev. A {\bf 65}, 043614 (2002).}
\bibitem{Rodrigues}
A. S. Rodrigues, P. G. Kevrekidis, R. Carretero-Gonz\`alez, D. J. Frantzeskakis, P. Schmelcher,
Phys. Rev. A {\bf 79}, 043603 (2009).


\bibitem{sakurai} J. J. Sakurai, {\it Modern Quantum Mechanics (Revised Edition)}, Addison Wesley, Reading, 1994, pp 158-174.



\bibitem{tf}S. Gautam and S. K. Adhikari, \pra  {\bf 92},  023616
  (2015).
\bibitem{fcv} S. Gautam and S. K. Adhikari, {\em Fractional-charge vortex in a spinor Bose-Einstein condensate},
arXiv:1601.01541.
\bibitem{Huepe}
 C. Huepe and M. E. Brachet, 
 Physica D {\bf 140}, 126 (2000);
 N. G. Berloff and P. H. Roberts, 
 J. Phys. A: Math. Gen. {\bf 33}, 4025 (2000);
 S. Rica, Physica D {\bf 148}, 221 (2001);
 C.-T. Pham, C. Nore, and M. E. Brachet, 
 Physica D {\bf 210}, 203 (2005);
 F. Pinsker and N. G. Berloff, 
 Phys. Rev. A {\bf 89}, 053605 (2014).
\bibitem{Watabe}
 S. Watabe, Y. Kato, and Y. Ohashi,
 Phys. Rev. A {\bf 84}, 013616 (2011).

\bibitem{BdG}
H. Pu and N. P. Bigelow,
Phys. Rev. Lett. {\bf 80}, 1134 (1998);
D. Gordon and C. M. Savage,
Phys. Rev. A {\bf 58}, 1440 (1998).
 \bibitem{num}S. Gautam and S. K. Adhikari, Phys. Rev A {\bf 90}, 043619 (2014);
 Phys. Rev. A {\bf 91}, 013624 (2015).  
 
\bibitem{cn}P. Muruganandam and S. K. Adhikari, Comput. Phys.
Commun. {\bf 180}, 1888 (2009); J. Phys. B {\bf 36}, 2501 (2003);
D. Vudragovic, I. Vidanovic,
A. Balaz, P. Muruganandam, and S. K. Adhikari,
Comput. Phys. Commun. {\bf 183}, 2021 (2012);
R. Kishor Kumar, L. E. Young-S., D. Vudragovi\'c, A. Balaz, P. Muruganandam, and  S.K. Adhikari,  Comput. Phys. Commun. {\bf 195}, 117  (2015). 
\bibitem{Inouye}
S. Inouye, M. R. Andrews, J. Stenger, H.-J. Miesner, D. M.
Stamper-Kurn, and W. Ketterle, Nature (London) {\bf 392}, 151
(1998); {C. Chin, R. Grimm, P. Julienne, and E. Tiesinga,
Rev. Mod. Phys. {\bf 82}, 1225 (2010).}

\bibitem{Hakim}
 V. Hakim,
 Phys. Rev. E {\bf 55}, 2835 (1997);
 N. Pavloff,
 Phys. Rev. A {\bf 66}, 013610 (2002).
\bibitem{Ji}
 A. C. Ji, W. M. Liu, J. L. Song, and F. Zhou, 
 Phys. Rev. Lett. {\bf 101}, 010402 (2008);
 M. Eto, K. Kasamatsu, M. Nitta, H. Takeuchi, and M. Tsubota, 
 Phys. Rev. A {\bf 83}, 063603 (2011);
 J. Lovegrove, M. O. Borgh, and J. Ruostekoski, 
 Phys. Rev. A {\bf 86}, 013613 (2012).

\end{thebibliography}
\end{document}